\documentclass[12pt]{iopart}
\usepackage{graphicx}

\usepackage{iopams}

\begin{document}

\title[Path dependent resistivity in Se doped CoS$_2$]{Path dependent resistivity study across field induced paramagnetic to ferromagnetic transition in Se doped CoS$_2$}

\author{Saroj Kumar Mishra and R Rawat\footnote{E-mail : rrawat@csr.res.in}}

\address{UGC-DAE Consortium for Scientific Research, University Campus, Khandwa Road, Indore-452001, India.}

\begin{abstract}
A systematic study of thermomagnetic irreversibility CoS$_{1.76}$Se$_{0.24}$ has been carried out. Our study shows that the resistivity at low temperature can be tuned by cooling in different magnetic fields and the critical field required for paramagnetic (PM) to ferromagnetic (FM) transition varies non-monotonically with temperature.  The field induced PM to FM transition results in giant positive magnetoresistance (MR) of about 160\% at 5 K. Measurements under CHUF (cooling and heating in unequal magnetic field) protocol show reentrant transition on warming under higher magnetic field (than that applied during cooling). It indicates that the glass like behaviour in this system can be explained in the framework of the kinetic arrest of first order transition.  Among the growing list of diverse system showing glass like arrested magnetic states, the present system is the first example where, kinetic arrest is observed for a disordered (here PM) to ordered (here FM) first order transition.
\end{abstract}

\pacs{75.30.Kz, 72.15.Gd, 75.60.Nt, 75.50.Cc}

\section{Introduction}
Transition metal di-chalcogenide, which crystallizes in simple cubic pyrite structure \cite{Elliott1960, Johnson1970}, show diverse magnetic and transport properties e.g. paramagnetic semiconductor to ferromagnetic metallic behaviour \cite{Jarrett1968}, superconductivity \cite{Bither1966}, metal-insulator transition\cite{Wilson1971, Mori1973}, quantum criticality \cite{Niklowitz2008}, giant magnetoresistance \cite{Yomo1979, Wada2014} etc. Among these, the compound CoS$_2$ is of interest due to itinerant electron metamagnetism \cite{Goto1997PRB}. It is reported to order ferromagnetically with transition temperature T$_{C}$=124 K, which is accompanied with increase in resistivity at T$_{C}$ \cite{Morris1967, Jarrett1968,  Adachi1969, Ogawa1971}. In contrast to ordinary ferromagnets, larger resistivity of FM state than PM state in CoS$_2$ is argued to be due to spin polarized state i.e. due to reduction in the density of state at Fermi level of minority spin band \cite{Ogawa1971, Wang2004}. With the application of pressure, T$_C$ is shifted to low temperature and appearance of thermal hysteresis indicates first order nature of the transition \cite{Elkin2014, Barakat2005,Goto1997PRB, Goto1997PhysicaB, Sato1969}. The nature of transition in CoS$_2$ under ambient pressure, itself has been a subject matter of debate \cite{Otero-Leal2008, Wang2004, Goto1997PRB, Barakat2005}. Recent work of Otero$-$Leal et al. suggest that it is weakly first order  and T$_C$ lies close to tricriticality \cite{Otero-Leal2008}. The transition vanishes around 5-6 GPa \cite{Barakat2005, Elkin2014} and under higher pressures the application of magnetic field result in itinerant metamagnetic behaviour \cite{Goto1997PRB}. Interestingly the effect of Se substitution for S, which increases the unit cell volume, is nearly same as that of pressure i.e. it also leads to decrease in T$_C$, which becomes first order and finally absence of transition for higher substitution of Se \cite{Wada2006, Wada2014, Adachi1981}. Such contradictory behaviour is argued to be resulting from stronger hybridization between transition metal d orbital and Se p orbitals in CoSe$_{2}$ \cite{Goto1997PRB}. Therefore competing AFM-FM interaction in Co(S$_{1-x}$Se$_{x}$)$_2$ decreases T$_C$ \cite{Wada2014, Adachi1981}.

	Though the composition with ferromagnetic ordering (i.e. $x<{0.12})$ has been studied extensively, there are few studies on system with $x\geq{0.12}$. Recently, Wada et al.\cite{Wada2014} studied field induced transition around x = 0.12 and observed giant positive magnetoresistance and magnetocaloric effect at low temperature. Their isothermal magnetoresistance data at 4.2 K for compositions x = 0.11 and 0.12 show an open loop (see figure 8 of \cite{Wada2014}) i.e. magnetoresistance is non-zero after field cycling. Similar non-zero MR can also be observed in the  MR data of Adachi et al.\cite{Adachi1981} for x = 0.12. The magnetization study in Co(S$_{1-x}$Se$_{x}$)$_2$ by Adachi et al. \cite{Adachi1970} in 1970 has also noted the presence of virgin curve lying outside the envelope curve in isothermal magnetization at 4.2 K (see figure 4 of \cite{Adachi1970}). To account for such behaviour, they suggested some kind of inhomogeneous stand point of view. NMR work by Panissod et al. \cite{Panissod1979} and Yasuoka et al. \cite{Yasuoka1979} showed coexistence of ferromagnetically ordered and nonmagnetic Cobalt. Latter transform to FM state with the application of magnetic field. The higher M at low field after field cycling  can be accounted by increased FM fraction at the expense of nonmagnetic fraction \cite{Panissod1979}. These thermomagnetic irreversibilities i.e. remanent MR after field cycling and virgin curve lying outside the envelope curve in isothermal M-H are similar to that observed in systems with kinetically arrested first order transition \cite{Roy2008, Chaddah2014} e.g. colossal magnetoresistance materials \cite{Kumar2006, Banerjee2006, Sathe2010, Rawat2013PRB}, intermetallics \cite{Manekar2001, Kushwaha2008, Rawat2013JPCM}, shape memory alloys \cite{Sharma2007, Banerjee2011, Siruguri2013}. The kinetic arrest of first order transition in these systems result in a glass like magnetic state at low temperatures.  In fact Sakata and Matsubara et al.\cite{Sakata1977, Matsubara1976} have proposed the possibility of glass like state due to random distribution of S and Se in Co(S$_{1-x}$Se$_{x}$)$_2$. However, the nature of such states in Se doped CoS$_2$ remains to be explored.

	To address these issues we have carried our systematic path dependent study of magnetic state in H-T space  using resistivity in CoS$_{1.76}$Se$_{0.24}$. This study brings out that the chracteristic of these thermomagnetic irreversibility can be associated with kinetic arrest of disordered broadened first order transition. Apart from demonstrating the tunability of PM-FM phase, the measurements under CHUF (cooling and heating under unequal magnetic field) protocol \cite{Banerjee2009} provide unambiguous evidence of glass like arrested PM phase.

\section{Experimental Details}

The polycrystalline sample with nominal composition CoS$_{1.76}$Se$_{0.24}$ is prepared by solid state reaction method. Constituent elements are mixed as per required stoichiometry and palletized after grinding. These pellets are sealed in quartz tube under a vacuum of $10^{-6}$ torr and heated to $250^0$C for 12hrs followed by at $400^0$C for 72hrs. These are pulverized again and above process is repeated with final annealing at $750^0$C for 72hrs. Powder x-ray diffraction measurement is carried out to check phase purity and lattice parameter. Absence of un-indexed peak indicates single phase nature of the sample and unit cell parameter is found to be 0.5578 nm, which is in good agreement with the existing literature \cite{Adachi1969, Krill1979, Johnson1970}. Resistivity measurements are carried out by standard four-probe technique using a homemade resistivity setup along with 8 T superconducting magnet system from Oxford Instruments, UK. All the in-field measurements are performed in longitudinal geometry and MR is defined as MR= ${{(\rho(H)- \rho(0))}/\rho(0)}$, where $\rho(0)$ is the resistivity in zero field and $\rho(H)$ is the resistivity in the presence of magnetic field (H).

\section{Results and discussion}

	Fig. 1[a] shows the temperature dependence of the electrical resistivity in the presence of various constant magnetic fields for CoS$_{1.76}Se_{0.24}$. For these measurements, labeled magnetic field is applied isothermally at 200 K and resistivity is measured during cooling (FCC) and subsequent warming (FCW). In the absence of applied magnetic field resistivity varies monotonically with temperature which indicates that it remains PM down to 5 K, which is consistent with existing literature \cite{Adachi1981, Wada2014}. For magnetic field $\geq$4 T, increase in resistivity with decrease in temperature indicates the presence of PM to FM transition. The broad transition can be attributed to quench disorder inherent in chemically substituted system, which leads to spatial distribution of transition temperature over sample volume on the length scale of correlation length \cite{Imry1979}. For 2 T magnetic field, though no distinct rise in resistivity is observed in FCC, the resistivity decreases with increase in temperature (FCW) and the presence of thermal hysteresis in resistivity is an indication of some transformation during cooling.

\begin{figure}[hbt]
		\begin{center}
			\includegraphics[width=7.5 cm]{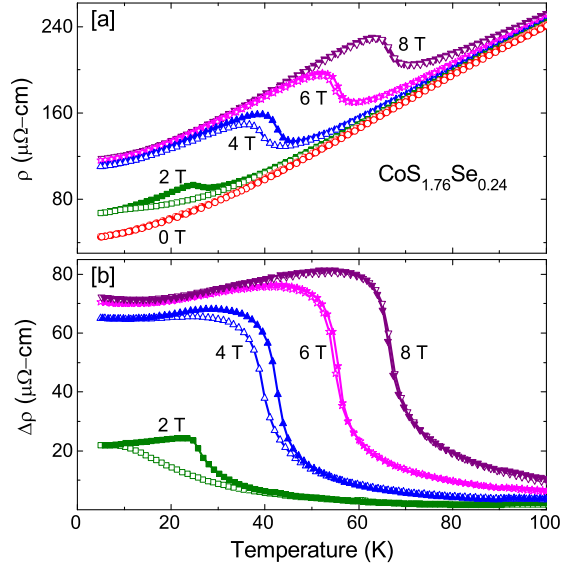}
		\end{center}
		\caption{\textbf{[a]} Resistivity($\rho$) versus temperature in presence of labelled magnetic field measured during cooling (open symbol) and subsequent warming (closed symbol). \textbf{[b]} Temperature dependence of resistivity difference ($\Delta \rho$) with respect to zero field resistivity (i.e. $\rho (H)- \rho (0)$).}
		\label{Figure1}
	\end{figure}

	Earlier resistivity studies in Co(S$_{1-x}$Se$_x$)$_2$ have shown that resistivity change associated with PM to FM transition remains nearly constant  \cite{Adachi1981,Wada2014}.	Our measurements (Fig. 1(a)) show that resistivity change associated with the magnetic transition is significantly smaller for 2 T magnetic field than that observed for higher field. To bring out this aspect, the difference of resistivity ($\Delta \rho$) with respect to zero field resistivity (i.e. $\rho (H)- \rho (0)$) is plotted as function of temperature in figure 1[b]. It shows that $\Delta \rho$ for 6 and 8 T magnetic field is about 70 $\mu\Omega$-cm at 5 K. However, for 2 T curve it remains around 20 $\mu\Omega$-cm. Considering that $\Delta \rho$ at 5 K is arising due to higher FM phase fraction, it can be inferred that cooling in 2 and 4 T magnetic field result in smaller FM phase fraction. Similar incomplete transition in resistivity were reported for doped CeFe$_{2}$\cite{Manekar2001, Roy2008}, Nd$_{0.5}$Sr$_{0.5}$MnO$_3$(NSMO)\cite{Rawat2007}, Co doped Mn$_{2}$Sb\cite{Kushwaha2008}, La-Pr-Ca-Mn-O (LPCMO) \cite{Sathe2010}, Pd doped FeRh\cite{Kushwaha2009}  etc. Particularly, the 2 T FCC and FCW curve appears to be similar to that observed for Pd doped FeRh in the presence of 6 T showing a broad gradual but incomplete transition in FCC and relatively sharp transition in FCW \cite{Kushwaha2009}. Such asymmetry in cooling and warming curves has also been seen in LPCMO thin film on mesoscopic length scale \cite{Rawat2013PRB}.
	
	Figure 2[a] to [c] shows isothermal MR measurement at various temperatures. For these measurements, sample is cooled in zero field from 200 K to measurement temperature and magnetic field is applied isothermally. Sharp resistivity increase with increase in magnetic field indicates PM to FM transition. It results in giant magnetoresistance, which reaches around 160\% at 1.5 K. Smaller MR value even after the completion of magnetic transition at high temperature can be attributed to increased phonon contribution to resistivity as the resistivity difference between PM and FM state is reported to remain almost constant with temperature \cite{Adachi1981, Wada2014}. At low temperature, as shown in figure 2[a], the forward curve (0 to 8 Tesla) shifts to left side whereas the return curve (8 to 0 Tesla) shifts to right side.  Whereas, at higher temperature (figure 2[b]-[c]) both forward and return curve moves to right side. Latter is expected for a conventional first order transition from low temperature high moment (here FM) state to high temperature low moment state (here PM). Similar opposing trend for forward and return curve has been shown in isothermal magnetoresistance measurement of  FeRh \cite{Kushwaha2009} and isothermal magnetization measurement of NSMO \cite{Rawat2007} and Ta doped HfFe$_2$ \cite{Rawat2013JPCM}. Another similarity which can be noticed between these and the present system is that presence of open loop in MR at low temperature i.e. MR does not become zero after the removal of applied magnetic field. This is highlighted in figure 2[a]. Similar open loop in isothermal MR measurement can also be observed in the MR data reported by Wada et al \cite{Wada2014} and Adachi et al. \cite{Adachi1981}.  The temperature dependence of remanent MR is shown as an inset in the figure 2[a], which shows that it vanishes above 20 K. The remanent MR at low temperature indicates that the magnetic state of the system is different before and after field cycling and a fraction of FM phase obtained with the application of 8 Tesla magnetic field is retained even after magnetic field removal. This fraction of remanent FM phase decreases with increase in temperature.
	
	\begin{figure}[t]
		\begin{center}
			\includegraphics[width=7.5 cm]{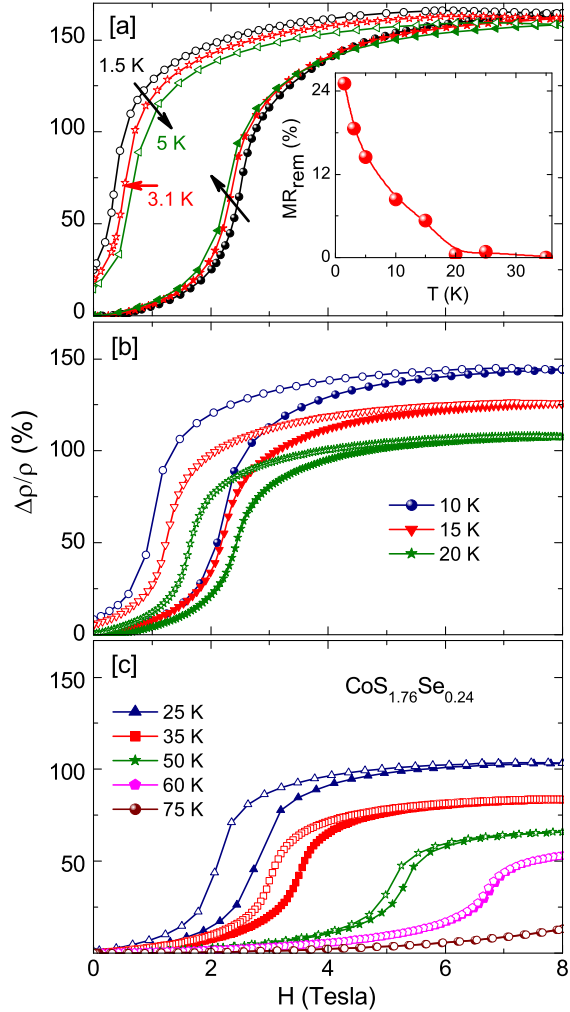}
		\end{center}
	\caption{\textbf{(a)-(c)}Isothermal magnetoresistance ($\Delta\rho/\rho$) vs H for CoS$_{1.76}$Se$_{0.24}$ at various temperature. Tilted arrows in the top panel indicate that forward (0$\rightarrow$ 8 T, solid symbol) and return curve (8 $\rightarrow$ 0 T, open symbol) shift in opposite direction with increase in temperature. Inset in the top panel shows the thermal variation of remanent MR (the MR at zero field after field cycling).}
		\label{Figure2}
	\end{figure}
	
	An open loop in MR (or different magnetic state after field cycling) can arise either for temperature lying within the zero field hysteresis region (supercooling/superheating effect) of $\rho-T$ curve or due to kinetic arrest of first order transition \cite{Kushwaha2008}. As shown in figure 3 of Kushwaha et al. \cite{Kushwaha2008}, the open loop due to supercooling and superheating is expected to arise within the zero field hysteresis region, but for only one direction of approach to measurement temperature. In the present case, as the low temperature state is FM, the slope of (H$_C$, T$_C$) line will be positive in H$-$T space and it corresponds to figure 3(b) of Kushwaha et al. \cite{Kushwaha2008}. For the measurement temperature lying within the hysteresis region of zero field resistivity, the magnetic state of the system before and after magnetic field cycling will be different if measurement temperature is reached by cooling, whereas it will remain unchanged if measurement temperature is reached by warming. Therefore for the present system an open loop in isothermal MR is expected when measurement temperature lying within the zero field hysteresis region is reached by cooling. However, no thermal hysteresis is observed in zero field resistivity. Even in the presence of 2 T magnetic field the $\Delta \rho$ curve for FCC shows that transition during cooling appears to be stopped at low temperature without completion (see figure 1). Secondly, magnitude of remanent MR is higher for lower temperature, which is expected for a kinetically arrested system \cite{Manekar2001, Rawat2007, Kushwaha2008}.
	
	Based on these isothermal MR measurements (shown in figure 2) a magnetic phase diagram is drawn, which is shown in figure 3. Here, the critical field for PM to FM transition (H$_{up}$) and FM to PM transition (H$_{dn}$) are taken as the magnetic field where d(MR)/dH shows maxima for forward and return curve, respectively. Transition temperature obtained from $\rho$-T measurement, for which transition temperature during cooling T* and warming T** are taken as the temperature of minima in the respective d$\rho$/dT curve, are also included in this figure. The obtained phase diagram is consistent with existing phase diagram, which report only T$_C$ (average of T* and T**) variation with H \cite{Adachi1979, Adachi1979b}. Present study shows the variation of hysteresis and brings out the non-monotonic variation of (H$_{up}$, T*) curve. Latter is shown to be anomalous in the case of NSMO, where it is observed for lower critical field (or transition during field decreasing cycle) \cite{Rawat2007}. There, such non-monotonic variation of critical field are explained considering the interplay of kinetic arrest band (below which first order transition is hindered) and supercooling band. Similar interplay of kinetic arrest band and supercooling band leads to non-monotonic variation of upper critical field in Ta doped HfFe$_2$ \cite{Rawat2013JPCM}, Gd$_5$Ge$_4$ \cite{Roy2007}, LPCMO \cite{Sharma2005, Wu2006} etc. In this picture, the low temperature behavior is dictated by the kinetic arrest band (which has negative slope) and the high temperature behavior is dictated by the supercooling band (which has a positive slope).
	
\begin{figure}[hbt]
		\begin{center}
			\includegraphics[width=7 cm]{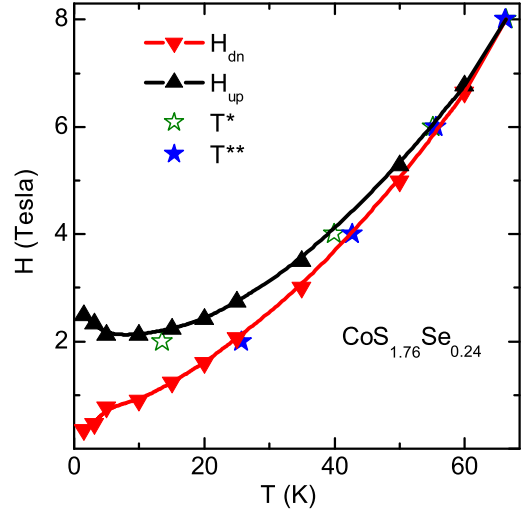}
		\end{center}
		\caption{H-T phase diagram for CoS$_{1.76}$Se$_{0.24}$ obtained from isothermal $\rho$-H (triangle) and $\rho$-T (star) measurements. It highlights non-monotonic variation of upper critical field.}
		\label{Figure3}
	\end{figure}

If the sample is cooled in a magnetic field lying between the overlapping region of kinetic arrest  and supercooling bands, it results in phase coexistence. The phase fraction can be tuned continuously from one end to other end by varying the magnetic field within this field window \cite{Banerjee2006}.  Results of such measurement in the present sample are shown in figure 4[a], where sample is cooled under the labeled magnetic field (H$_{an}$) from 70 K to 1.5 K and magnetic field is then isothermally changed from H$_{an}$ to 2.5 T. From this figure it is evident that for H$_{an} >$2.5 T resistivity is a function of H$_{an}$ at 1.5 K and 2.5 T, and it remains almost constant with magnetic field reduction from H$_{an}$ to 2.5 T. Whereas, for H$_{an}$$\leq$ 2.5 T resistivity curve merges to each other on isothermal increase of magnetic field from H$_{an}$ to 2.5 T. These observations show that the FM phase fraction increases with increase in cooling field and this phase fraction remains invariant on isothermal reduction of magnetic field to 2.5 T at 1.5 K.

\begin{figure}[t]
		\begin{center}
		\includegraphics[width=7.5 cm]{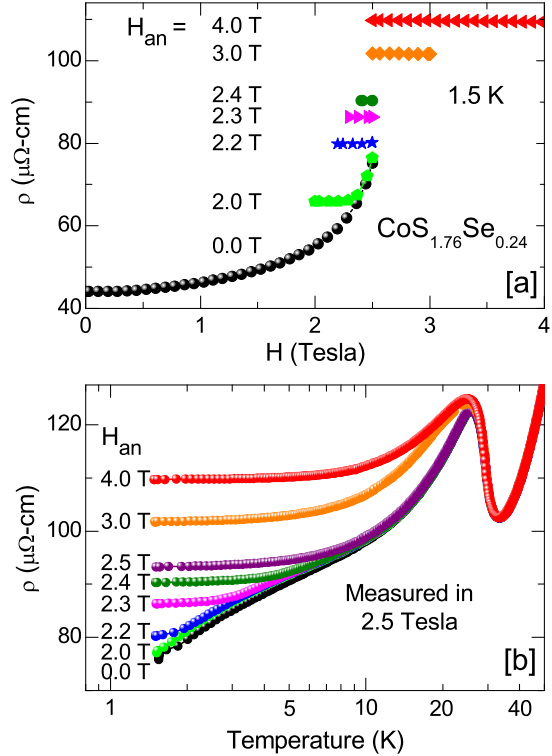}
		\end{center}
		\caption{\textbf{(a)} $\rho$ vs. H for H$_{an}$ to 2.5 Tesla at 1.5 K and then \textbf{(b)} $\rho$ vs. T is measured during warming in the presence of 2.5 Tesla. For these measurements sample is cooled to 1.5 K in the presence of magnetic field H$_{an}$.}
		\label{Figure4}
	\end{figure}

	Which of these states is equilibrium state is verified using CHUF protocol \cite{Banerjee2009}. Under this protocol, sample is cooled in a various magnetic field H$_{an}$ and measurement is carried out during warming in a magnetic field H$_{w}$. Depending on the observation of reentrant transition and sign of H$_w$-H$_{an}$ equilibrium state is determined. Figure 4[b] shows one such measurement in the present system. All the curves in this figure correspond to resistivity measurement during warming in the presence of 2.5 T magnetic field. Difference between these curves lies in the cooling history i.e. H$_{an}$ under which sample is cooled to 1.5 K. For H$_{an}>$2.5 T only one transition is observed i.e. around 30 K. Whereas, for H$_{an}<$2.5 T, it shows a reentrant transition. Sharper resistivity increase on warming at low temperature indicates PM to FM transition, which is not expected for a conventional supercooled state. It indicates that low field state i.e. PM state is the non-equilibrium state. Therefore these measurements show that with lowering temperature the PM to FM transition is kinetically arrested and the observation of transition to FM state on warming in 2.5 Tesla is akin to devitrification of glass which is followed by FM to PM transition at higher temperature akin to melting. As expected, the devitrification temperature depends on the cooling history. Also the curves for lower H$_{an}$ merge together before merging with the curve for higher H$_{an}$. Such topology of devitrification curve is expected for anticorrelated kinetic arrest and supercooling band i.e. regions with lower kinetic arrest temperature have higher supercooling temperature \cite{Kumar2006}. Barring few exception \cite{Rawat2012}, anticorrelation between kinetic arrest band and supercooling band has been a common feature for kinetically arrested systems so far.

\section{Conclusions}
	To conclude, we studied the path dependence of resistivity in CoS$_{1.76}$Se$_{0.24}$. The field induced transition results in giant positive magnetoresistance of about 160\% at 5 K. Our systematic study of thermomagnetic irreversibility reveals remanent MR, opposite trend in forward and return curves of isothermal MR measurement and non-monotonic variation of upper critical field at low temperatures. These features indicate presence of kinetic arrest of first order PM-FM transition. The interplay of supercooling and kinetic arrest band results in tunable PM-FM phase fraction at low temperature. The CHUF measurements reveal reentrant transition during warming when sample is cooled in lower magnetic field. 	In the light of present work, it can be concluded that the glass like magnetic state in Co(S$_{1-x}$Se$_x$)$_2$ discussed by Sakata and Matsubara et al. \cite{Sakata1977, Matsubara1976} can be described in the frame work of kinetic arrest of first order transition. It is to be noted here that so far kinetic arrest has been observed for two ordered state separated by the first order transition. Whereas, in the present system the first order transition occurs from a disordered PM state to ordered FM state.

\section{Acknowledgement}
	Layanta Behera, Mukul Gupta are acknowledged for XRD measurements. P. Chaddah is acknowledged for fruitful suggestions.

{}


\section{References}
\begin{thebibliography}{}

\bibitem[1]{Elliott1960} N. Elliott, J Chem. Phys. \textbf{33}, 903 (1960).
\bibitem[2] {Johnson1970} V. Johnson and A. Wold, J. Solid State Chem. \textbf{2}, 209 (1970).
\bibitem[3] {Jarrett1968} H. S. Jarrett, W. H. Cloud, R. J. Bouchard, S. R. Butler, C. G. Frederick, and J. L. Gillson, Phys. Rev. Lett. \textbf{21}, 617 (1968).
\bibitem[4] {Bither1966} T. A. Bither, C. T. Prewitt, J. L.Gillson, P. E. Bierstedt, R. B. Flippen, and H. S. Young, Solid State Communications \textbf{4}, 533 (1966).
\bibitem[5] {Wilson1971} J. A. Wilson and G. D. Pitt, Philosophical Magazine \textbf{23}, 1297 (1971).
\bibitem[6] {Mori1973} N. Mori, T. Mitsui, and S. Yomo, Solid State Communications \textbf{13}, 1083 (1973).
\bibitem[7] {Niklowitz2008} P. Niklowitz, P. L. Alireza, M. J. Steiner, G. G. Lonzarich, D. Braithwaite, G. Knebel, J. Flouquet and J. A. Wilson, Phys. Rev. B \textbf{77}, 115135 (2008).
\bibitem[8] {Yomo1979} S. Yomo, J. Phys. Soc. Jpn \textbf{47}, 1486 (1979).
\bibitem[9] {Wada2014} H. Wada, D. Kawasaki, and Y. Maekawa, IEEE Trans. Magn. \textbf{50}, 2501806 (2014).
\bibitem[10] {Goto1997PRB} T. Goto, Y. Shindo, H. Takahashi, and S. Ogawa, Phys. Rev. B \textbf{56}, 14019 (1997).
\bibitem[11] {Morris1967} B. Morris, V. Johnson, and A. Wold, J. Phys. Chem. Solids \textbf{28}, 1565 (1967).
\bibitem[12] {Adachi1969} K. Adachi, K. Sato, and M. Takeda, J. Phys. Soc. Jpn. \textbf{26}, 631 (1969).
\bibitem[13] {Ogawa1971} S. Ogawa and T. Teranishi, Phys. Lett. A \textbf{36}, 407 (1971).
\bibitem[14] {Wang2004} L. Wang, T. Y. Chen, and C. Leighton, Phys. Rev. B \textbf{69}, 094412 (2004).
\bibitem[15] {Elkin2014} F. Elkin, I. Zibrov, A. Novikov, S. S. Khasanov, V. A. Sidorov, A. E. Petrova, T. A. Lograsso, J. D. Thompson, and S. M. Stishov, Solid State Commun. \textbf{181}, 41 (2014).
\bibitem[16] {Barakat2005} S. Barakat, D. Braithwaite, P. Alireza, K. Grube, M. Uhlarz, J. Wilson, C. Pfleiderer, J. Flouquet, and G. Lonzarich, Physica B \textbf{359-361}, 1216 (2005).
\bibitem[17] {Goto1997PhysicaB} T. Goto, Y. Shindo, S. Ogawa, and T. Harada, Physica B \textbf{237-238}, 482 (1997).
\bibitem[18] {Sato1969} K. Sato, K. Adachi, T. Okamoto and E. Tatsumoto, J. Phys. Soc. Jpn. \textbf{26}, 639 (1969).
\bibitem[19] {Otero-Leal2008} M. Otero-Leal, F. Rivadulla, M. Garcia-Hernandez, A. Pineiro, V. Pardo, D. Baldomir and J. Rivas, Phys. Rev. B \textbf{78}, 180415 (2008).
\bibitem[20] {Wada2006} H. Wada, A. Mitsuda, and K. Tanaka, Phys. Rev. B \textbf{74}, 214407 (2006).
\bibitem[21] {Adachi1981} K. Adachi, M. Matsui, and Y. Omata, J. Phys. Soc. Jpn. \textbf{50}, 83 (1981).
\bibitem[22] {Adachi1970} K. Adachi, K. Sato, and M. Matsuura, J. Phys. Soc. Jpn. \textbf{29}, 323 (1970).
\bibitem[23] {Panissod1979} P. Panissod, G. Krill, M. Lahrichi, and M. F. Lapierre-Ravet, J. Phys. C: Solid State Phys. \textbf{12}, 4281 (1979).
\bibitem[24] {Yasuoka1979} H. Yasuoka, N. Inoue, M. Matsui, and K. Adachi, J. Phys. Soc. Jpn. \textbf{46}, 689 (1979).
\bibitem[25] {Roy2008} S. B. Roy, P. Chaddah, V. K. Percharsky, and K. A. Gschneidner, Jr. Acta Materials \textbf{56}, 5895 (2008).
\bibitem[26] {Chaddah2014} P. Chaddah (2014), arXiv-1410.3254.
\bibitem[27] {Kumar2006} K. Kumar, A. K. Pramanik, A. Banerjee, P. Chaddah, S. B. Roy, S. Park, C. L. Zhang, and S. -W. Cheong, Phys. Rev. B \textbf{73}, 184435 (2006).
\bibitem[28] {Banerjee2006} A. Banerjee, K. Mukherjee, K. Kumar, and P. Chaddah, Phys. Rev. B. \textbf{74}, 224445 (2006).
\bibitem[29] {Sathe2010} V. G. Sathe, A. Ahlawat, R. Rawat, and P. Chaddah, J. Phys.: Condens. Matter \textbf{22}, 176002 (2010).
\bibitem[30] {Rawat2013PRB} R. Rawat, P. Kushwaha, D. K. Mishra, and V. G. Sathe, Phys. Rev. B \textbf{87}, 064412 (2013).
\bibitem[31] {Manekar2001} M. A. Manekar, S. Chaudhary, M. K. Chattopadhyay, K. J. Singh, S. B. Roy, and P. Chaddah, Phys. Rev. B. \textbf{64}, 104416 (2001).
\bibitem[32] {Kushwaha2008} P. Kushwaha, R. Rawat, and P. Chaddah, J. Phys.: Condens. Matter. \textbf{20}, 022204 (2008).
\bibitem[33] {Rawat2013JPCM} R. Rawat, P. Chaddah, P. Bag, P. D. Babu, and V. Siruguri, J. Phys.: Condens. Matter. \textbf{25}, 066011 (2013).
\bibitem[34] {Sharma2007} V. K. Sharma, M. K. Chattopadhyay, and S. B. Roy, Phys. Rev. B \textbf{76}, 140401 (2007).
\bibitem[35] {Banerjee2011} A. Banerjee, P. Chaddah, S. Dash, K. Kumar, A. Lakhani, X. Chen, and R. V. Ramanujan, Phys. Rev. B \textbf{84}, 214420 (2011).
\bibitem[36] {Siruguri2013} V. Siruguri, P. D. Babu, S. D. Kaushik, A. Biswas, S. K. Sarkar, M. Krishnan, and P. Chaddah, J. Phys.: Condens. Matter. \textbf{25}, 496011 (2013).
\bibitem[37] {Sakata1977} M. Sakata, F. Matsubara, Y. Abe, and S. Katsura, J. Phys. C: Solid State Phys. \textbf{10}, 2887 (1977).
\bibitem[38] {Matsubara1976} F. Matsubara and M. Sakata, Prog. Theo. Phys. \textbf{55}, 672 (1976).
\bibitem[39] {Banerjee2009} A. Banerjee, K. Kumar, and P. Chaddah, J. Phys.: Condens. Matter. \textbf{21}, 026002 (2009).
\bibitem[40] {Krill1979} G. Krill, P. Panissod, M. Lahrichi, and M. F. Lapierre-Ravet, J. Phys. C: Solid State Phys. \textbf{12}, 4269 (1979).
\bibitem[41] {Imry1979} Y. Imry and M. Wortis, Phys. Rev. B \textbf{19}, 3580 (1979).
\bibitem[42] {Rawat2007} R. Rawat, K. Mukherjee, K. Kumar, A. Banerjee, and P. Chaddah, J. Phys.: Condens. Matter. \textbf{19}, 256211 (2007).
\bibitem[43] {Kushwaha2009} P. Kushwaha, A. Lakhani, R. Rawat, and P. Chaddah, Phys. Rev. B. \textbf{80}, 174413 (2009).
\bibitem[44] {Adachi1979} K. Adachi, M. Matsui, and M. Kawai, J. Phys. Soc. Jpn. \textbf{46}, 1474 (1979).
\bibitem[45] {Adachi1979b} K. Adachi, M. Matsui, Y. Omata, H.Mollymoto, M.Motokawa, and M. Date, J. Phys. Soc. Jpn. \textbf{47}, 675 (1979).
\bibitem[46] {Roy2007} S. B. Roy, M. K. Chattopadhyay, A. Banerjee, P. Chaddah, J. D. Moore, G. K. Perkins, L. F. Cohen, K. A. Gschneidner, Jr., and V. K. Pecharsky, Phys. Rev. B. \textbf{75}, 184410 (2007).
\bibitem[47] {Sharma2005} P. A. Sharma, S. B. Kim, T. Y. Koo, S. Guha, and S.-W. Cheong, Phys. Rev. B \textbf{71}, 224416 (2005).
\bibitem[48] {Wu2006} W. Wu, C. Israel, N. Hur, S. Park, S. -W. Cheong, and A. de Lozanne, Nature Mater. \textbf{5}, 881 (2006).
\bibitem[49] {Rawat2012} R. Rawat, P. Chaddah, P. Bag, K. Das, and I. Das, J. Phys.: Condens. Matter. \textbf{24}, 416001 (2012).
\end{thebibliography}
\end{document}